\documentstyle [12pt]{article}
\advance\hoffset by -.5 in
\advance\voffset by -1 in
\begin{document}
\def\pz{\p_z}
\def\a{\alpha}
\def\b{\beta}
\def\g{\gamma}
\def\d{\delta}
\def\e{\epsilon}
\def\x{\xi}
\def\f{\phi}
\def\j{\psi}
\def\t{\tau}
\def\L{\Lambda}
\def\l{\lambda}
\def\p{\partial}
\def\o{\omega}
\def\O{\Omega}
\def\z{\zeta}
\def\Tau{{\rm T}}
\def\cA{{\cal A}}
\def\tcA{\tilde{\cal A}}
\def\cK{{\cal K}}
\def\tf{\tilde{f}}
\def\tg{\tilde{g}}
\def\th{\tilde{h}}
\def\rarr{\rightarrow}
\def\mp{\mapsto}
\def\hL{\hat{L}}
\def\res{{\rm res}}
\def\tr{{\rm tr}}
\def\tres{\tr~\res}
\def\cF{{\cal F}}
\def\et{\eta}
\def\cV{{\cal V}}
\def\s{\sum}
\def\cL{{\cal L}}
\def\cH{{\cal H}}
\def\hw{\hat{w}}
\def\hv{\hat{v}}
\def\hhw{\hat{\hw}}
\def\hhv{\hat{\hv}}
\def\exi{\exp\sum_{l\neq k}\xi_l}
\def\emxi{\exp(-\sum_{l\neq k}\xi_l)}
\def\bg{{\bf g}}
\def\ov{\overline}
\def\un{\underline}
\def\+{_{(+)}}
\def\-{_{(-)}}

\begin{center}
\begin{bf}
\begin{Large}
On $\tau$-functions of Zakharov-Shabat and other\\ matrix hierarchies
of integrable equations \\
\end{Large}

\vspace{.5in}
L. A. Dickey\\
\end{bf}
\vspace{.3in}
University of Oklahoma, Norman, OK 73019\footnote
{e-mail: ldickey@nsfuvax.math.uoknor.edu}
\end{center}

\vspace{.2in}
\begin{abstract}
Matrix hierarchies are: multi-component KP, general Zakharov-Shabat (ZS) and
its special cases, e.g., AKNS. The ZS comprises all integrable systems having a
form of zero-curvature equations with rational dependence of matrices on a
spectral parameter. The notion of a $\tau$-function is introduced here in the
most general case along
with formulas linking $\tau$-functions with wave Baker functions. The method
originally invented by Sato et al. for the KP hierarchy is used. This method
goes immediately from definitions and does not require any assumption about the
character of a solution, being the most general. Applied to the matrix
hierarchies, it involves considerable sophistication. The paper is
self-contained and does not expect any special prerequisite from a reader.
\end{abstract}

{\bf 1. Introduction.}\\

Integrable systems of differential equations exist not isolated but
united in large communities called hierarchies. All equations inside a
hierarchy are commuting with each other. The first known hierarchies were
generalized Korteweg-de Vries (KdV) hierarchies, one for every natural number
$n$. (For detail, see, e.g., [9]). Then an immense Kadomtsev-Petviashvili (KP)
hierarchy was found which united all the KdV's. Those hierarchies consisted of
scalar equations. Almost immediately they were generalized to matrix equations.
They formed ``multi-component" KdV's and KP.

All the above mentioned hierarchies are
generated by linear differential (KdV) or pseudo-differential (KP) operators of
arbitrary orders. Equations of another type are generated by matrix first order
differential operators linearly depending on a spectral parameter. These are
AKNS (for Ablowitz, Kaupp, Newell and Segur) with $2\times 2$ matrix first
order
operators, they were generalized by Dubrovin to $n\times n$ matrices; we call
the latter AKNS-D hierarchies. The next generalization is when linear operators
depend on a parameter as polynomials of any degree. Finally, the most general
case involves arbitrary rational dependence on a parameter. These equations are
called general Zakharov-Shabat (ZS) equations. They also form a hierarchy (see
[7]). The hierarchy with polynomial dependence on a parameter is a special case
of the general ZS when there is a single pole, at infinity; we call this
hierarchy s-p ZS. All KdV's and AKNS's are nothing but reductions of the
general ZS hierarchy. The exact definitions will be given below.

The importance of the theory of integrable system was essentialy enhanced with
the invention of the ``tau"-function by mathematicians of the
Kyoto school, see [1], [2]. This one single function of infinitely many ``time"
variables replaces infinitely many dynamical variables, coefficients of linear
differential or pseudo-differential operators. It happened that this
function linked integrable systems to Lie algebras representations and to
many problems of modern physics, such as conformal field theory, matrix models
in the statistical physics,
2-dimensional gravity and string theory. Up to now, all these achievements
applied solely to scalar hierarchies, $n$th KdV and KP. There are also some
published results about the multi-component KP. Concerning the general ZS
eqations and their $\tau$-functions, we know the only work [10] done in
very abstract terms; it is difficult to extract concrete formulas from it.
The aim of the present article is to
fill up this gap. Physicists have not turned yet their attention to the general
ZS hierarchy (except some particular equations of these hierarchy). We believe
that its time will come sooner or later.\\

The paper is self-contained and, formally speaking, does not require a special
prerequisite (see also [9]). It was easier not to start with the most
complicated case of the general ZS but to pass gradually from the simplest to
the most difficult model referring when needed to what was proven before.

In the first part of the paper we deal with the
multi-component KP (mcKP). It is defined in [1]. In [3] and [4] there are
formulas written for its $\tau$-function, in both the
articles without proofs. Therefore it is difficult to guess what was the way
they followed. Most probably, they used the techniques of free fermion
representations. Meanwhile, those authors had suggested their own excellent
method which was invented in [2] for KP, based on nothing but the bilinear
identity, i.e., being close to very first definitions.
The advantage of this approach is its full generality, independence of the
origin and the nature of a solution. Our first goal was to adjust this
method to the mcKP hierarchy. Basically, the method remains
the same as in [2], however, it becomes a little tricky. (In [5] we
derived the $\tau$-function in terms of the Grassmannian, in [6] found it for
special, algebraic geometrical solutions; in contrast to that, we discuss now
the general case).

The next part is devoted to the simplest special case of the ZS hierarchy,
namely, the single-pole hierarchy (s-p ZS). It is closely connected with the
mcKP since it is proven below that a Baker function of the s-p ZS is at the
same time that of mcKP, and the s-p ZS is a subhierarchy of the mcKP. A similar
statement was made before in [5].

Then we introduce a ``not normalized" s-p ZS hierarchy which differs from the
previous one by the fact that the expansion of its Baker function in powers of
a
spectral parameter starts with a matrix of a general form, not from the unity.
It can be reduced to the normalized ZS. Nevertheless, it is convenient to study
this case separately because it provides a good preparation for the general ZS
where one cannot normalize Baker functions simultaneously at all poles.

Finally, and this is the main point, we treat the general ZS
hierarchy. It is discussed in [7] in what sense one can understand the
totality of all ZS equations as a hierarchy, i.e., as a set of commuting vector
fields. There are definitions of a Baker function, of a corresponding
Grassmannian, etc. in that paper. However, it is lacking a concept of the
$\tau$-function. We are doing this now. The main results of the present
article are contained in the theorems of sect. 4 and sect. 7 and the
proposition 3 and its corollary of sect. 5.

Despite the absence of a general definition and of a proof of the
existence of the $\tau$-function, there were a few examples of that function
found earlier. In [7] this is done for soliton-type solutions and quite
recently, in [8], for algebraic geometrical solutions that can be expressed in
terms of $\theta$-functions. Those examples were stimulating for the present
study.\\

{\bf 2. Multi-component KP.}\\

Let $$ L=A\partial +u_0+u_1\partial^{-1}+\cdots,~~\p=d/dx$$ be a
pseudo-differential operator
where $u_i$ are $n\times n$ matrices, $A={\rm diag}(a_1,...,a_n),\:\:a_i$ are
distinct non-zero constants. Diagonal elements of $u_0$ are assumed to be zero.

Let $R_{\alpha}=\sum_{j=0}^{\infty}R_{j\alpha}\partial^{-j}$, $\alpha=1,...,n,$
where $R_{0\alpha}=E_{\alpha}$, $E_{\alpha}$ is a matrix having only one
non-zero element on the $(\a,\a)$ place which is equal to 1; $R_{\a}$ is
supposed to satisfy $$ [L,R_{\alpha}]=0.$$ It is shown below that such matrices
exist being $$R_{\alpha}R_{\beta}=\delta_{\alpha
\beta}R_{\alpha},\hspace{.1in}\sum_{\a=1}^n R_{\alpha}=I$$ (i.e. this is a
spectral decomposition of the unity). The mcKP hierarchy (multi-component KP)
is
$$\p_{k\alpha}L=[(L^kR_{\alpha})_+,L],~\p_{k\a}R_{\b}=[(L^kR_{
\alpha})_+,R_{\b}]~~\p_{k\a}=\p/\p t_{k\a},~~k=0,1,...;~\a=
1,...,n $$ and $t_{k\a}$ are the ``time variables" of the hierarchy. The
subscript + refers, as usual, to a purely differential part of a
pseudo-differential operator, $(\sum a_k\p^k)_+=\sum_{k\geq 0} a_k\p^k$, $A_-=A
-A_+$.

It can be shown that the equations for different $k,\a$ commute. The variables
$x$ and $t_{k
\alpha}$ are not independent: $$\partial=\sum_{\alpha}a_{\alpha}^{-1}\partial_{
1\alpha}$$ (Greek indices always run from 1 to $n$).

Let $$L=\hw A\p\hw^{-1},~~{\rm where}~~\hw=\hw(A\p)=\sum_0^{\infty}w
_i(A\p)^{-i},~w_0=I;$$ Then $R_{\alpha}=\hw E_{\alpha}\hw^{-1}$ has all needed
properties. Put
$$w=\hw(A\p)\exp\xi(t,z)=\hw(z)\exp\xi(t,z);~~{\rm where}~~\xi(t,z)=\sum_{k=
0}^{\infty}\sum_{\alpha=1}^nz^kE_{\a}t_{k\alpha}.$$
This is the {\em Baker function}; it satisfies the equations
$$Lw=zw,\hspace{.1in}{\rm
and}\hspace{.1in}\partial_{k\alpha}w=(L^kR_{\alpha})_+
w. $$ The latter equation is equivalent to
$$ \partial_{k\alpha}\hw=-(L^kR_{\alpha})_-\hw.$$

{\bf Remark 1.} It is very important to note that the series $\hw$ are defined
up to a multiplication on the
right by series $\sum_0^\infty c_i\p^{-i}$ with constant diagonal matrices
$c_i$ where $c_0=I$. Correspondingly, the Baker function is defined up to a
multiplication by $\sum_0^\infty a_iz^{-i}$. Two functions which differ by
such a factor are said to be equivalent. For two equivalent Baker functions the
Lax operator $L$ is the same. All the formulas below will be obtained up to the
equivalence.

We have $\p_{0\a}\hw=-(R_\a-E_\a)\hw=-\hw E_a+E_\a\hw=
[E_\a,\hw].$ Symmetries related to ``zero" time variables $t_{0\a}$ are
similarity transformations with constant matrices.

The {\em adjoint Baker function} is $$w^a=(\hw^*(A\p))^{-1}\exp(-\xi(t,z))
$$ where the star means the conjugation: for every matrix $X$ the equality
$(X\p)^*=-\p X^*$ holds where $X^*$ is the transpose of $X$.

The equations $$L^*w^a=zw^a,\hspace{.1in}{\rm and}\hspace{.1in}
\partial_{k\alpha}w^a=-(L^kR_{\alpha})_+^*w^a $$ hold.      \\

{\bf Remark 2.} Our definition of the mcKP differs from that in [1],[2] and [3]
where $u_0=0$ and $A=I$. It is easy to show that in our definition the
coefficients of the equations are local in terms of $u_i$'s, i.e.,
differential polynomials of them. Indeed, the dressing formula $L=\hw A\p\hw^{-
1}$ permits to express every differential polynomial in elements of $w_i$'s as
a differential polynomial in elements of $u_i$'s which is also an ordinary
polynomial in $w_i$'s (i.e., it does not depend on derivatives of $w_i$'s).
Then, the elements of $R_\a$'s are such polynomials, too. Let us show that, in
fact, they do not depend on $w_i$'s at all. Let us give to $\hw$ an
infinitesimal deformation $\d\hw$ such that $L$ is not changed. This means
that $\d L=[\d\hw\cdot\hw^{-1},A\p]=0.$ This easily implies that the matrix
$K=\d\hw\cdot\hw^{-1}$ is constant and diagonal. Now, $\d R_\a=[K,E_\a]=0.$
The rest is clear. The fact that all diagonal elements of $A$ are distinct
is crucial. It is easy to compute that otherwise $R_\a$ are not local.
If one is only interested in the hierarchy in terms of Baker functions, not
of the operator $L$, then this distinction is not important.

The significance of the mcKP, as well as KP, is in their universality. \\

{\bf Proposition. Universality of the mcKP hierarchy.} {\sl If an expression of
the form $$w=\hw(A\p)\exp\xi(t,z)=\hw(z)\exp\xi(t,z);~~{\rm where}~~\hw(A\p)=
\sum_0^{\infty}w_i(A\p)^{-i},~w_0=I$$ satisfies arbitrary equations
$\p_{k\a}w=\ov{B}_{k\a}w$ with some differential operators $\ov{B}_{k\a}$ then
this is nothing but mcKP.}\\

Indeed, the given equations yield
$$0=\p_{k\a}\hw\cdot e^\xi+\hw E_\a z^ke^\xi-\ov{B}_{k\a}w=\p_{k\a}\hw\cdot e^
\xi+\hw E_\a(A\p)^ke^\xi-\ov{B}_{k\a}w.$$ Letting $L=\hw A\p\hw^{-1}$ and $R_\a
=\hw E_\a\hw^{-1}$ we have $\p_{k\a}\hw\cdot\hw^{-1}+R_\a L^k-\ov{B}_{k\a}=0.$
Taking the positive part of this equation, we get $\ov{B}_{k\a}=(R_\a L^k)_+$
and the negative part is $\p_{k\a}\hw=-(R_\a L^k)_-\hw$. This is the equation
of the hierarchy. $\Box$\\

{\bf 3. Bilinear identity.}\\

The so-called bilinear identity is basic for Sato's theory. \\

{\bf Lemma.} {\sl Let $\Phi=\sum\Phi_i(A\p)^i$ and $\Psi=\sum\Psi_i(A\p)^i$ be
two pseudo-differential operators ($\Psi{\rm DO}$). Then the equality
$$\res_\p\Phi\Psi^*=\res_z(\Phi e^\xi)A^{-1}(\Psi e^{-\xi})^*$$ holds.}\\

The notations res$_\p$ and res$_z$ mean, as usual, coefficients of $\p^{-1}$
and $z^{-1}$.

{\em Proof.} It is easy to check that both the left- and the right-hand side
are equal to $\sum\Phi_iA^{-1}\Psi_{-i-1}^*(-1)^{i+1}$. $\Box$      \\

{\bf Proposition.} {\sl If $\Phi=\sum\Phi_i(A\p)^i$ is a $\Psi$DO, and
$w=\Phi\exp\xi(t,z)$, $w^a=(\Phi^*)^{-1}\exp(-\xi(t,z))$ then $$\res_z(\p^iw)A
^{-1}(w^a)^*=0.\eqno{(1a)}$$ Moreover, if $w$ depends on infinitely many
variables $t_{i\a}$, $i=0,1,...,\a=1,...,n$ and satisfies a system of
differential equations of the form $\p_{i\a}w=B_{i\a}w$ where $B_{i\a}$ are
any differential (matrix) operators in $\p=\sum a_\a^{-1}\p_{1\a}$ then
$$ \res_z(\p_{i_1\a_1}\p_{i_2\a_2}...\p_{i_s\a_s}w)A^{-1}(w^a)^*=0\eqno{(1b)}$$
for an arbitrary set of indices $i_1,\a_1,i_2,\a_2,...,i_s,\a_s$. This happens,
e.g., when $w$ is a Baker function of the mcKP hierarchy.

Conversely, if there are two expressions of the form $w=\sum_0^\infty w_i(t,z)z
^{-i}\exp\xi$ and \break $w^a=\sum v_i(t,z)z^{-i}\exp(-\xi)$ with $w_0=v_0=I$,
and Eq.(1a) holds for them, then letting $\Phi=\sum w_i(A\p)^{-i}$ we will have
$w=\Phi\exp\xi$ and $w^a=(\Phi^*)^{-1}\exp(-\xi)$.

Moreover, if the stronger equality (1b) holds, then $w$ and $w^a$ are the Baker
and the adjoint Baker functions of the mcKP.}\\

{\em Proof.} We have $$\res_z(\p^iw)A^{-1}(w^a)^*=\res_z(\p^i
\Phi e^\xi)A^{-1}((\Phi^*)^{-1}e^{-\xi})^*$$
$$=\res_\p\p^i\Phi((\Phi^*)^{-1})^*
=\res_\p\p^i\Phi\Phi^{-1}=\res_\p\p^i=0.$$ This proves the first statement.
Now,
the equations $\p_{i\a}w=B_{i\a}w$ allow to express all the derivatives
$\p_{i\a}$ in terms of $\p$ and then to apply (1a) which proves (1b).

The first statement of the converse proposition can be obtained in the
following way. Let $\Phi=\sum w_i(A\p)^{-i}$ and $\Psi=\sum v_i(-A\p)^{-i}$
then $w=\Phi\exp\xi$ and $w^a=\Psi\exp(-\xi)$. We have $$0=\res_z(\p^i\Phi
e^\xi
)A^{-1}(\Psi e^{-\xi})^*=\res_\p\p^i\Phi\Psi^*$$ for all $i\geq 0$. The
operator $\Phi\Psi^*$ is $I+O(\p^{-1})$, and the last equality implies that the
negative part is zero. Hence $\Phi\Psi^*=I$ and $\Psi=(\Phi^*)^{-1}$.

Now, let (1b) hold. Put $L=\Phi A\p\Phi^{-1}$. We have $$((\p_{k\a}\Phi)+(L^kR^
\a)_-\Phi)e^\xi=(\p_{k\a}\cdot\Phi-\Phi(A\p)^kE_\a+(L^kR^\a
)_-\Phi)e^\xi=(\p_{k\a}-(L^kR^\a)_+)\Phi e^\xi.$$ Then, applying the
assumption and the lemma, $$0=\res_z\p^i(\p_{k\a}-(L^kR^\a)_+)wA^{-1}(w^a)^*=
\res_z\p^i(\p_{k\a}-(L^kR^\a)_+)\Phi e^\xi A^{-1}((\Phi^*)^{-1}e^{-\xi})^*$$
$$=\res_\p\p^i((\p_{k\a}\Phi)+(L^kR^\a)_-\Phi)\Phi^{-1}.$$ This yields
$(\p_{k\a}\Phi)+(L^kR^\a)_-\Phi=0$, i.e., the equation of the hierarchy. $\Box$

The bilinear identity can also be written in a dual form
$$\res_z wA^{-1}(\p_{i_1\a_1}\p_{i_2\a_2}...\p_{i_s\a_s}w^a)^*=0.$$
The proof is similar.

Very often they use the identity in the form $$\res_zw(t,z)A^{-1}(w^a(t',z))^*=
0$$ where $t'$ is another set of values $t_{k\a}$. This identity makes sense
as a formal expansion in powers of $t'_{k\a}-t_{k\a}$.\\

{\bf 4. $\tau$-function.}\\

Let $G_\a(\z)$ be an operator of translation acting as $$G_\a(\z)
f(t,z)=f(...,t_{k\g}-\d_{\a\g}{1\over k\z^k},...,z).$$ Let $$N_\a(\z)=-\sum_
{j=0}^\infty\z^{-j-1}\p_{j\a}+\p_\z,~~\p_\z=\p/\p\z.$$ It is easy to see
that $N_\a(\z)G(\z)f(t,z)=0$.

According to the bilinear identity,
$$\res_zw(t,z)A^{-1}G_\b(\z)(w^a(t,z))^*=0.$$ We have
$$G_\b(\z)\exp(-\sum_{k\g}
t_{k\g}E_\g z^k)=(I-E_\b+(1-{z\over\z})^{-1}E_\b)\exp(-\sum_{k\g}t_{k\g}E_\g z^
k)$$ as it is easy to check. If $w(z)=\hw(z)\exp\xi$ and $w^a(z)=\hw^a(z)\exp(-
\xi)$ then $$\res_z\hw(z)A^{-1}(I-E_\b+(1-{z\over\z})^{-1}E_\b)G_\b(\z)(\hw^a
(z))^*=0.$$ It is easy to see that if $f(z)=\sum f_iz^i$ then $\res_z
f(z)(1-z/\z)^{-1}=\z f_-(\z)$ where the subscript ``$-$" symbolizes the
negative
part of the series. We have $\hw=I+w_1z^{-1}+...$ and, as a simple calculation
shows, $(\hw^a)^*=I-Aw_1A^{-1}z^{-1}+...$ . The identity becomes
$$w_1(I-E_\b)A^{-1}-(I-E_\b)G_\b w_1A^{-1}+\z[\hw(\z)A^{-1}E_\b G_\b(\z)(\hw^a(
\z))^*]_-=0.$$The $(\b,\b)$th element of this matrix identity is $\hw_{\b\b}(\z
)a_\b^{-1}G_\b(\z)(\hw^a(\z))_{\b\b}^*-a_\b^{-1}I=0.$ Thus, we have $$\hw_{\b\b
}(\z)G_\b(\z)(\hw^a(\z))_{\b\b}=I.\eqno{(2)}$$ The shifted $(\hw^a(\z))_{\b\b}$
happens to be just the inverse of $\hw_{\b\b}(\z)$.

Let us take now the $(\a,\b)$th element of the matrix identity:
$$-a_\b^{-1}G_\b(\z)w_{1,\a\b}+\z\hw_{\a\b}(\z)a_\b^{-1}G_\b(\z)(\hw^a(\z))_{\b
\b}=0.$$ Using (2), transform this to $$G_\b(\z)w_{1,\a\b}=\z\hw_{\a\b}(\z)
(\hw_{\b\b}(\z))^{-1}.\eqno{(3)}$$

Now, consider a more complicated relation which also follows from the bilinear
identity: $$\res_z w(z)A^{-1}G_\a(\z_1)G_\b(\z_2)(w^a(z))^*=0.$$ In the case
when $\a=\b$ this reduces to $$\res_z\hw(z)A^{-1}[I-E_\b+(1-\frac z\z_1)^{-1}
(1-\frac z\z_2)^{-1}E_\b]G_\b(\z_1)G_\b(\z_2)(\hw^a(z))^*=0.$$ Taking the
$(\b,\b)$th element we have $$\res_z\hw_{\b\b}(z)(1-\frac z\z_1)^{-1}(1-\frac z
\z_2)^{-1}G_\b(\z_1)G_\b(\z_2)(\hw^a(z))_{\b\b}=0$$ or $$\res_z\hw_{\b\b}(z)[
\z_1^{-1}(1-\frac z\z_1)^{-1}-\z_2^{-1}(1-\frac z\z_2)^{-1}]G_\b(\z_1)G_\b(\z_2
)(\hw^a(z))_{\b\b}=0$$ which yields $$\hw_{\b\b}(\z_1)G_\b(\z_1)G_\b(\z_2)(
\hw^a(\z_1))^*=
\hw_{\b\b}(\z_2)G_\b(\z_1)G_\b(\z_2)(\hw^a(\z_2))_{\b\b}.$$ Using (2), we
obtain
$${G_\b(\z_2)\hw_{\b\b}(\z_1)\over\hw_{\b\b}(\z_1)}={G_\b(\z_1)\hw_{\b\b}(\z_2)
\over\hw_{\b\b}(\z_2)}.$$ Taking a logarithm and denoting $\ln\hw=f$ we get
$$(G_\b(\z_2)-1)f_{\b\b}(\z_1)=(G_\b(z_1)-1)f_{\b\b}(\z_2).\eqno{(4)}$$

In the case when $\a\neq\b$ the identity is $$\res_z\hw(z)A^{-1}[I-E_\a-E_\b+
(1-\frac z\z_1)^{-1}E_\a+(1-\frac z\z_2)^{-1}E_\b]G_\a(\z_1)G_\b(\z_2)(\hw^a(z)
)^*=0.$$ The $(\a,\a)$th element of this matrix identity is
$$\z_1\hw_{\a\a}(\z_1)a_\a^{-1}G_\a(\z_1)G_\b(\z_2)(\hw^a(\z_1))_{\a\a}+\z_2
\hw_{\a\b}(\z_2)a_\b^{-1}G_\a(\z_1)G_\b(\z_2)(\hw^a(\z_2))_{\b\a}^*-\z_1a_
\a^{-1}I=0.$$  The $(\b,\a)$th element is
$$\z_2\hw_{\b\b}(\z_2)a_\b^{-1}G_\a(\z_1)G_\b(\z_2)(\hw^a(\z_2))_{\b\a}^*+\z_1
\hw_{\b\a}(\z_1)a_\a^{-1}G_\a(\z_1)G_\b(\z_2)(\hw^a(\z_1))_{\a\a}=0.$$
Eliminating $(\hw^a)_{\b\a}^*$ from two equations and applying (2), we obtain
$$-\hw_{\b\b}(\z_2)+(\hw_{\a\a}(\z_1)\hw_{\b\b}(\z_2)-\hw_{\b\a}(\z_1)\hw_{\a\b
}(\z_2))G_\b(\z_2)(\hw_{\a\a}(\z_1))^{-1}=0.$$ Take a logarithm:
$$\ln\hw_{\b\b}(\z_2)=\ln(\hw_{\a\a}(\z_1)\hw_{\b\b}(\z_2)-\hw_{\b\a
}(\z_1)\hw_{\a\b}(\z_2))-G_\b(\z_2)\ln\hw_{\a\a}(\z_1)$$ and subtract this
equation from one obtained by permutation of $\a$ and $\b$, $\z_1$ and $\z_2$.
The result is $$(G_\b(\z_2)-1)f_{\a\a}(\z_1)=(G_\a(\z_1)-1)f_{\b\b}(\z_2),~~
f=\ln\hw.\eqno{(5)}$$ Eq.(4) is a special case of this one when $\a=\b$.

Now we have to prove the existence of a function $\tau(t)$ such that
$f_{\a\a}(\z)=(G_\a(\z)-1)\ln\tau$. If the operator $(G_\a(\z)-1)$ had an
inverse, this would immediately follow from (5). This operator has a kernel
consisting of constants (with respect to $\{t_{k\a}\}$). Let us apply the
operator $N_\a(\z_1)$ to Eq.(5): $$G_\b(\z_2)N_\a(\z_1)f_{\a\a}(\z_1)-
N_\a(\z_1)f_{\a\a}(\z_1)=\sum_{j=0}^\infty \z_1^{-j-1}\p_{j\a}f_{\b\b}(\z_2).$$
Then multiply this by $\z_1^i$ and take res$_{\z_1}$: $$b_{i\a}\equiv\res_{\z_
1}\z_1^iN_\a(\z_1)f_{\a\a}(\z_1)=G_\b(\z_2)\res_{\z_1}\z_1^iN_\a(\z_1)f_{\a\a}(
\z_1)+\p_{i\a}f_{\b\b}(\z_2),$$ i.e., $$b_{i\a}=G_\b(\z_2)b_{i\a}
+\p_{i\a}f_{\b\b}(\z_2).\eqno{(6)}$$Here $(i,\a)$ is an arbitrary pair of
indices, one can replace them by $(j,\g)$: $$b_{j\g}=G_\b(\z_2)b_{j\g}
+\p_{j\g}f_{\b\b}(\z_2).$$ Differentiating the first equality with respect to
$t_{j\g}$, the second with respect to $t_{i\a}$ and subtracting, we have
$(G_\b(\z_2)-1)(\p_{j\g}b_{i\a}-\p_{i\a}b_{j\g})=0$
whence $\p_{j\g}b_{i\a}-\p_{i\a}b_{j\g}$ is
a constant. It is not difficult to see from the definition of $b_{i\a}$ that
this constant can be only zero. Thus, $\p_{j\g}b_{i\a}=\p_{i\a}b_{j\g}$. This
implies the existence of a function of the variables $\{t_{i\a}\}$, we call it
ln $\tau(t)$, such that $b_{i\a}=\p_{i\a}\ln\tau$: $$\res_{\z}\z^i(-\sum_0
^\infty z^{-j-1}\p_{j\a}+\p_\z)\ln\hw_{\a\a}(\z)=\p_{i\a}\ln\tau.\eqno{(7)}$$
The equation (6) yields that $\p_{i\a}f_{\b\b}(\z)=(G_\b(\z)-1)b_{i\a}=
(G_\b(\z)-1)\p_{i\a}\ln\tau$ and $f_{\b\b}(\z)=(G_\b(\z)-1)\ln\tau+$const. In
more detail, this formula looks like this: $$\hw_{\b\b}(\z)=c_\b(\z){\tau(...,t
_{k\g}-\d_{\b\g}\cdot1/(k\z^k),...)\over\tau(t)}.\eqno{(8)}$$ In the numerator
only the variables $t_{k\g}$ with $\g=\a$ are shifted. The constant $c_\b(\z)$
is a series $c_\b(\z)=\sum_{i=0}^\infty c_{i\b}z^{-i}$ with $c_{0\b}=1$.

We have obtained this formula only for diagonal elements of $\hw$ yet. Eq.(7)
is
a conversion of Eq.(8). Let $C=$diag $c_\b(\z)$, a constant diagonal matrix.
Then the Baker function $wC^{-1}$ is equivalent to $w$. For this function (8)
holds with $c_\b=1$.

Let us return to Eq.(3). We find $\hw_{\a\b}(\z)=\z^{-1}G_\b(\z)w_{1,\a\b}
\cdot\hw_{\b\b}$, substituting $\hw_{\b\b}$ from (8) and denoting
$$\tau_{\a\b}(t)=\tau(t)w_{1,\a\b},~~\a\neq\b,\eqno{(9)}$$ this becomes $\hw_{
\a\b}(\z)=\z^{-1}G_\b(\z)\tau_{\a\b}\cdot(\tau(t))^{-1}$, or $$\hw_{\a\b}(\z)=
\z^{-1}c_\b(\z){\tau_{\a\b}(...,t_{k\g}-\d_{\b\g}\cdot1/(k\z^k),...)
\over\tau(t)},~~\a\neq\b.\eqno{(10)}$$ Thus, only those variables $t_{k,\g}$
are shifted whose index $\g$ coincides with the number of the column, $\b$.
Thus, we have a theorem:\\

{\bf Theorem.} {\sl For any Baker function there are functions $\tau(t)$ and
$\tau_{\a\b}(t)$ and constant series $c_\b(\z)$ such that Eqs.(8) and (10)
hold. Coefficients $c_\b(\z)$ are insignificant if a Baker functions is
considered to within the equivalence.}\\

The formulas (8) and (10) are the main formulas of the theory of the
$\tau$-function.   \\

{\bf Remark.} The definition of the $\tau$-function and the derivation of the
formulas (8) and (10) based on the bilinear identity does not depend on the
property of the matrix $A$ to have distinct elements on the diagonal. It
remains valid even if $A=I$. This will be used in the next section.\\

{\bf 5. Single-pole Zakharov-Shabat hierarchy.}\\

The general Zakharov-Shabat
equation is $[I\p+U(z),I\p_t+V(z)]=0$ where matrices $U$ and $V$ are rational
functions of a parameter $z$. In [7] it was explained in what sense the
totality of all possible equations of this form can be considered as one
hierarchy. Now we are interested in the case when both the functions $U(z)$
and $V(z)$ have a single pole which is at infinity, i.e., they are polynomials
in $z$.

Let $\hw=\sum_{i=0}^\infty w_iz^{-i}$ be a formal series, $w_0=I$.\\

{\bf Definition.} The single-pole ZS hierarchy is the totality of all the
equations $$\p_{l\a}\hw=-(z^lR_\a)_-\hw~~{\rm where}~~R_\a=\hw E_\a\hw^{-1}.
\eqno{(11)}$$
The subscript ``$-$" refers to the negative part of an expansion in powers of
$z$.

Letting $w=\hw\exp\xi(t,z)$, where $\xi$ is as before, we get
an equivalent form of the equations of the hierarchy $$\p_{l\a}w=B_{l\a}w,~~
B_{l\a}=(z^lR_\a)_+.\eqno{(12)}$$ The same equation can also be expressed
as $$w\p_{l\a}\cdot
w^{-1}=\hw(\p_{l\a}-z^lE_\a)\hw^{-1}=I\p_{l\a}-B_{l\a}.\eqno
{(13)}$$ Thus, dressing of $I\p_{l\a}$ yields a first-order differential
operator (13) that is a $l$th degree polynomial in $z$.
The expression $w$ is called a {\em formal Baker function}.

It can be proven that the operators $\p_{l\a}$ commute. This fact and Eq.(13)
imply that operators $I\p_{l\a}-B_{l\a}$ commute, i.e.,$$\p_{l\a}
B_{m\b}-\p_{m\b}B_{l\a}-[B_{l\a},B_{m\b}]=0.\eqno{(14)}$$

Let $\l_l,~l=0,...,m+1$ be a sequence of constant diagonal matrices,
$\l_l=$diag
$(\l_{l\a})$, $\l_{m+1,\a}=a_\a$ being distinct,
 and $\p=-\sum_{l=0}^{m+1}\sum_{\a=1}^n\l_{l\a}\p_{l\a}$. Set
$$L=-\sum_{l=0}^{m+1}\sum_{\a=1}^n\l_{l\a}(I\p_{l\a}-B_{l\a})=I\p+U\eqno{(15)}$$
where $U=\sum_{i=0}^{m+1}\sum_{\a=1}^n\l_{i\a}B_{l\a}$. Then
$$L=w\p w^{-1}=I\p+U=I\p+u_0+u_1z+...+u_mz^m-Az^{m+1},~~A={\rm diag}~a_\a.
\eqno{(16)}$$

The hierarchy equations imply $$\p_{m\b}L=[B_{m\b},L].$$
If $M$ is another operator defined in the same way as $L$ with other matrix
diagonal coefficients, $\mu_{l\a}$ instead of $\l_{l\a}$, then $[L,M]=0$. This
is exactly the ZS equation with a single pole.

The notion of the {\em equivalence} is the same as for the mcKP: two Baker
functions are equivalent if they differ by a factor on the right which is a
constant diagonal matrix series. Then $B_{l\a}$'s remain the same along with
all
differential operators $L$.\\

{\bf Proposition 1. Universal property.} {\sl Let $\hw$ be a series $\hw=\sum_
{i_0}^\infty w_iz^{-i}$, $w_0=I$ and $w=\hw\exp\xi$. All the functions depend
on variables $t_{k\a}$. If $w$ satisfies an equation of the form $\p_{k\a}w
=\overline{B}_{k\a}w$ where $\overline{B}_{k\a}w$ is a polynomial in $z$ then
this is an equation of the hierarchy, i.e. $\overline{B}_{k\a}=(z^kR_\a)_+$.}
\\

{\em Proof.} We have
$$0=\p_{k\a}\hw\cdot e^\xi+\hw E_\a z^ke^\xi-\overline{B}_{k\a}w $$ and
$$0=\p_{k\a}\hw\cdot \hw^{-1}+\hw E_\a z^k\hw^{-1}-\overline{B}_{k\a}.$$
Taking the positive part, we obtain $\overline{B}_{k\a}=(\hw E_\a
z^k\hw^{-1})_+=(z^kR_\a)_+$. $\Box$\\

{\bf Proposition 2.} {\sl Let $\hw$ be a series $\hw=\sum_{i_0}^\infty w_iz^{-i
}$, $w_0=I$ and $w=\hw\exp\xi$. All the functions depend on variables
$t_{k\a}$.
Then if $w$ satisfies the hierarchy equations (12) then the following
bilinear identity $$\res_z z^i\p_{k_1\a_1}...\p_{k_s\a_s}w\cdot w^{-1}=0
\eqno{(17)}$$ holds for arbitrary sets of indices, $i\geq 0$.

Conversely, if there is another series $\hv=\sum_{i_0}^\infty v_iz^{-i}$,
$v_0=I$, $v=\exp(-\xi)\hv$ and $$\res_z z^i\p_{k_1\a_1}...\p_{k_s\a_s}w
\cdot v=0$$ for all sets of indices then $v=w^{-1}$ and $w$ is a Baker function
of the hierarchy.}\\

{\em Proof}. Let $w$ be a Baker function of the hierarchy. Then, by virtue of
the equation (12), the left-hand side of (12) is a residue of a polynomial
which is zero.

Conversely, $\res_zz^iwv=0$ for all $i$ implies that $(wv)_-=0$, $\hw\hv=I$,
and $v=w^{-1}$. We have further
$$(\p_{k\a}\hw+(z^kR_\a)_-\hw)e^\xi=(\p_{k\a}-(z^kR_\a)_+)w$$ where $R_\a$
is defined as in (11). Using the assumption, one gets $$0=\res_zz^i(\p_{k\a}-
(z^kR_\a)_+)w\cdot w^{-1}=\res_zz^i(\p_{k\a}\hw+(z^kR_\a)_-\hw)\hw^{-1}.$$ This
implies $\p_{k\a}\hw+(z^kR_\a)_-\hw=0$ which is the hierarchy equation (11).
$\Box$\\

{\bf Proposition 3.} {\sl Baker functions of the mcKP are those of s-p ZS,
more than that, the action of the operators $\p_{k\a}$ on them is
the same in both the hierarchies. In other words, the s-p ZS hierarchy is
a restriction of the mcKP.}\\

{\em Proof.} The first statement follows from the fact that a Baker
function of the s-p ZS hierarchy satisfies a bilinear identity (17) stronger
than (1b). The converse part of the proposition of Sect.3 can be applied
(letting $A^{-1}(w^a)^*=w^{-1}$). It also
follows from the second statement. Let us prove the latter.
Let $A$ be an arbitrary constant diagonal matrix with distinct
diagonal elements. Put $\p=\sum_\a a_\a^{-1}\p_{1\a}.$ Let $w$ be a Baker
function of the single-pole ZS hierarchy. Then $w$ satisfies Eq.(17) for every
multi-index. It suffices to show that $w$ satisfies the set of equations of the
form $\p_{k\a}w=\ov{B}_{k\a}w$ for all $k$ and $\a$ where $\ov{B}_{k\a}$ are
differential operators in $\p$ (see Proposition, Sect.2). We have obvious
relations:
\begin{eqnarray*}
\p_{k\a}w&=&(E_\a z^k+O(z^{k-1}))e^\xi,  \\
A^q\p^qw&=&(z^q+O(z^{q-1}))e^\xi.
\end{eqnarray*}
whence $\p_{k\a}w-E_\a A^k\p^kw=O(z^{k-1})\exp\xi=(V_{k-1}z^{k-1}+O(z^{k-2}))
\exp\xi$. The process can be prolonged: $\p_{k\a}w-E_\a A^k\p^kw-V_{k-1}A^{k-1}
\p^{k-1}w=O(z^{k-2})\exp\xi$ etc. In the end we have $$\p_{k\a}w-\ov{B}_{k\a}w
\equiv\p_{k\a}w-E_\a A^k\p^kw-V_{k-1}A^{k-1}\p^{k-1}w-...-V_0w=O(z^{-1})e^\xi
$$
where $\ov{B}_{k\a}$ is a differential operator. Now, the bilinear identity $$
\res_zz^i(\p_{k\a}-\ov{B}_{k\a})w\cdot w^{-1}=0$$ where $(\p_{k\a}-\ov{B}_{k\a}
)w\cdot w^{-1}=O(z^{-1})$ implies that $(\p_{k\a}-\ov{B}_{k\a})w\cdot w^{-1}=0
$, and $(\p_{k\a}-\ov{B}_{k\a})w=0$ as required. $\Box$   \\

{\bf Remark.} It can seem strange that $\ov{B}_{k\a}w$ whose elements are
differential polynomials with respect to $\p$ coincides with $B_{k\a}$ where
only ordinary polynomials are involved. The explanation is that one of
equations of the s-p ZS hierarchy is $\p w\equiv\sum a_\a^{-1}\p_{1\a}w=
B_{1\a}w$ or, in detail, $(\p+[A^{-1}, w_1]-A^{-1}z)w=0$. It enables us to
eliminate all the derivatives. It is spectacular and instructive (though
cumbersome) to verify the statement of the last proposition directly even in
the simplest case of $\p_{2k}$.\\

{\bf Corollary.} {\sl The $\tau$-functions for the s-p ZS hierarchy exist
and they are a special case of those for the mcKP.}\\

%
%
%
%

{\bf 6. Not normalized s-p ZS hierarchy.} \\

The hierarchy in the last section was
normalized, in the sense that $w_0=I$. Now $w_0$ also will depend on time
variables, $w_0(t)$. The definition of the hierarchy (11) must be adjusted to
this requirement since (11) implies that $w_0=$const.

Let $A\+$ symbolize the purely positive part of an expansion in powers of $z$,
i.e., without the constant term, and $A\-$ negative part with the constant
term,
i.e. the constant term passes from the positive part to the negative one.
Eq.(11) will be replaced by $$\p_{l\a}\hw=-(z^lR_\a)\-\hw,~~R_\a=\hw E_\a\hw^{-
1}\eqno{(18)}$$ and Eq.(12) by $$\p_{l\a}w=B_{l\a}=(z^lR_a)\+.$$ It can be
proven that the operators $\p_{l\a}$ commute as well.\\

{\bf Proposition 1.} {\sl If $\hw$ satisfies (18) then $\hv=w_0^{-1}\hw$
satisfies (11).}\\

{\em Proof.} Eq.(18) implies $$\p_{l\a}w_0=-(z^l\hw E_\a\hw^{-1})_0w_0
=-w_0(z^l\hv E_\a\hv^{-1})_0.\eqno{(19)}$$ Then
$$ \p_{l\a}\hv=-w_0^{-1}\p_{l\a}w_0\cdot w_0^{-1}\hw+w_0^{-1}\p_{l\a}\hw=
$$ $$=(z^l\hv E_\a\hv^{-1})_0\hv-w_0^{-1}(z^l\hw E_\a\hw^{-1})\-\hw=
(z^l\hv E_\a\hv^{-1})_0\hv-(z^l\hv E_\a\hv^{-1})\-\hv$$ $$=-(z^l\hv E_\a\hv^{-1
})_-\hv.$$This is exactly Eq.(11). $\Box$

The proposition 1 allows to express $\hv$ in terms of a $\tau$ function.
However, this is not what we need, $w_0$ remains indefinite. In order to
determine it one has to solve the linear equation (19). We show further
that the whole function $\hw$ has an expression in terms of $\tau$-functions
$$\hw_{\a\b}(\z)={\tau_{\a\b}(...,t_{k\g}-\d_{\b\g}\cdot1/(k\z^k),...)
\over\tau(t)}\eqno{(20)}$$ for both $\a=\b$ and $\a\neq\b$.

First of all, one must write the bilinear identity. In this case, Eq.(17)
holds in a stronger form: if $s>0$ then (17) holds for $i\geq -1$, if $s=0$
then it holds for $i\geq 0$ while res$_zz^{-1}w\cdot w^{-1}=I$. The identity
can be written in the dual form where all the derivatives act on $w^{-1}$
rather than on $w$.\\

{\bf Proposition 2.} {\sl To every Baker function there exist functions $\tau$
and $\tau_{\a\b}$ such that Eq.(20) holds up to an equivalence, i.e.,
$$\hw_{\a\b}(\z)=c_\b(\z){\tau_{\a\b}(...,t_{k\g}-\d_{\b\g}\cdot1/(k\z^k),...)
\over\tau(t)}$$ where $c_{\b}(z)$ are constant series.}\\

It follows from the bilinear identity (17) that $$\res_zz^iw\cdot G_\b(\z)
w^{-1}=\left\{\begin{array}{ll}0,~&{\rm if}~i\geq 0\\I,~&{\rm if}~i=-1\end
{array}\right.$$ As in sect. 4 this transforms to
$$\res_zz^i\hw(z)(I-E_\b+(1-{z
\over\z})^{-1}E_\b)G_\b(\z)\hw^{-1}=\left\{\begin{array}{ll}0,~&{\rm if}~i\geq
0\\I,~&{\rm if}~i=-1\end{array}\right.$$  We have $$\res_zz^i\hw(z)(I-E_\b)G_\b
(\z)\hw^{-1}(z)$$ $$+\z(z^i\hw(z)G_\b(\z)E_\b\hw^{-1}(z))_-|_{z=\z}=\left\{
\begin{array}{ll}0,~&{\rm if}~i\geq 0\\I,~&{\rm if}~i=-1\end{array}\right.$$
Let $i=-1$. Then this equality becomes
$$w_0(I-E_\b)G_\b(\z)w_0^{-1}+\hw(\z)G_\b(\z)E_\b\hw^{-1}(\z)=I$$ or,
multiplying by $G_\b w_0$, $$w_0(I-E_\b)+\hw(\z)G_\b(\z)E_\b\hw^{-1}(\z)w_0=
G_\b(\z)w_0.$$ For the $(\a,\b)$th element this is $$\hw_{\a\b}G_\b(\hv^{-1})_
{\b\b}=G_\b(w_0)_{\a\b}.\eqno{(21)}$$ It is easy to see that the case $i=0$
gives the same for $\hv$ as it was in sect. 4 for $\hw$; in particular, the
analogues of (2) and (3), and the possibility to express $\hv$ in terms of a
$\tau$-function. Eq.(2) becomes $$G_\b(\hv^{-1})_{\b\b}=(\hv_{\b\b})^{-1}.\eqno
{(22)}$$ We even do not need to prove this since we knew this in advance. We
know also that there is a function $\tau(t)$ and constant series $c_\b$ such
that $\hv_{\b\b}=c_\b G_\b\tau\cdot\tau^{-1}$. With the help of Eq.(22), Eq.(21
) transforms to $$\hw_{\a\b}(\hv_{\b\b})^{-1}=G_\b(w_0)_{\a\b}.\eqno{(23)}$$

Now, let $$\tau_{\a\b}=\tau (w_0)_{\a\b}.\eqno{(24)}$$ Using (23) and (22), we
have $${G_\b\tau_{\a\b}\over\tau}={G_\b\tau\over\tau}\cdot G_\b(w_0)_{\a\b}=
c_\b^{-1}\hv_{\b\b}\hw_{\a\b}(\hv_{\b\b})^{-1}=c_\b^{-1}\hw_{\a\b}$$ as
required. $\Box$

Notice, that Eqs.(22) and (23) look very nice being put together in the form
\begin{eqnarray*}
G_\b(\hv^{-1})_{\b\b}&=&(\hv_{\b\b})^{-1},\\
G_\b(\hw\hv^{-1})_{\a\b}&=&\hw_{\a\b}(\hv_{\b\b})^{-1}.
\end{eqnarray*}

{\bf 7. The general ZS hierarchy.} \\

This hierarchy was introduced in [7]. Let $a_k$, $k=1,...,m$ be a given set of
complex numbers. Let, for every $k$, $$\hw_k=\sum_0^
\infty w_{ki}(z-a_k)^i,$$ be a formal series. The entries of $n\times n$
matrices $w_{ki}$, $w_{ki,\a\b}$ are just letters. We consider the algebra $\cA
_w$ of polynomials of all this entries and $(\det w_{k0})^{-1}$. The formal
series $\hw_k$ can be inverted within this algebra. Let $$R_{k\a}=\hw_kE_{\a}
\hw_k^{-1};~~R_{k\a l}=R_{k\a}(z-a_k)^{-l}$$ where $E_\a$ is, as before, a
matrix with only one non-vanishing element, equal 1, on the $(\a,\a)$ place.

We have the following objects. Such quantities as $\hw_k$ and $R_{k\a l}$
are formal series, or jets, at the points $a_k$. The algebra of all such jets
will be called $J_k$ and $J=\oplus J_k$. If $j_k\in J_k$ is a
jet then $j_k^-$ symbolizes its principal part, i.e., a sum of negative
powers of $z-a_k$, and $j_k^+$ the rest of the series. Correspondingly, the jet
algebras split into parts, $J_k=J_k^+\oplus J_k^-$. If the principal part
contains finite number of terms (and we tacitly assume this unless the opposite
is said or is evident from a context) it can be considered as a global
meromorphic function; the algebra of global meromorphic functions is $G$. A
global function gives rise to a jet at every $a_k$. In particular, $j_k^-$ can
be considered as a jet at a point $a_{k_1}$, different from $a_k$, more
precisely, as an element of $J_{k_1}^+$. And finally, there will be formal
products of jets or of global functions by expressions of the form $\exp\xi_k$
where $$\xi_k=\sum_{\a=1}^n\sum_{l=0}^\infty t_{k\a l}E_\a(z-a_k)^{-l}.$$

{\bf Definitions. (i)} {\sl A hierarchy corresponding to a fixed set $\{a_k\}$
is the totality of equations
$$\p_{k\a l}\hw_{k_1}=\left\{\begin{array}{l}-R_{k\a l}^+\hw_{k_1},~~k=k_1\\
{}~~R_{k\a l}^-\hw_{k_1},~~{\rm otherwise}\end{array}\right.,~~~\p_{k\a
l}=\p/\p t_{k\a l}.\eqno{(25)}$$ In the second case $R_{k\a l}^-$ is
considered as an element of $J_{k_1}^+$, see above; $t_{k\a l}$ are some
variables.}

{\bf (ii)} {\sl A ZS hierarchy is an inductive limit of hierarchies with fixed
sets $\{a_k\}$, with respect to a natural embedding of a hierarchy
corresponding
to a subset into a hierarchy corresponding to a larger set, as a subhierarchy.}
\\

In this article we deal with the hierarchy corresponding to a fixed set $\{a_k
\}$. There was proven in [7] that all the equations of the hierarchy commute.
The following proposition readily can be
checked by a simple straightforward computation:\\

{\bf Proposition 1.} {\sl A dressing formula $$ \hw_{k_1}(\p_{k\a l}-E_{\a}
(z-a_k)^{-l}\d_{kk_1})\hw_{k_1}^{-1}=\p_{k\a l}-B_{k\a l},~~B_{k\a l}=
R_{k\a l}^-\eqno{(26)}$$ is equivalent to Eq.(25).}\\

The operator $\p_{k\a l}-B_{k\a l}$ is assumed to act in $J_{k_1}$. However, it
does not depend on $k_1$ at all and can be considered as a global function of
$z$ with the only pole of the $l$th order at $a_k$.

Let $$w_k=\hw_k\exp\xi_k.$$

{\bf Definition.} {\sl The collection $w=\{w_k\}$ is the formal Baker function
of the hierarchy.} \\

Eq.(25) can be written in terms of the Baker function as
$$\p_{k\a l}w_{k_1}=B_{k\a l}w_{k_1}\eqno{(27)}$$ and Eq.(26) as $$w_{k_1}\p_{k
\a l}w_{k_1}^{-1}=\p_{k\a l}-B_{k\a l}.\eqno{(28)}$$

{\bf Proposition 2.} {\sl All the operators $\p_{k\a l}-B_{k\a l}$ commute.}\\

{\em Proof.} This is a corollary of the fact that $\p_{k\a l}$ commute and Eq.
(28). $\Box$\\

One can consider arbitrary linear combinations of the above constructed
operators,$$ L=\sum_{k,\a,l}\l_{k\a l}(\p_{k\a l}-B_{k\a l})=\p+U$$
where $\p=\sum_{k,\a,l}\l_{k\a l}\p_{k\a l}$ and $U=-\sum_{k,\a,l}\l_{k\a l}B_
{k\a l}$. Two such operators commute which yields equations of the
Zakharov-Shabat type $$\p U_1-\p_1 U=[U_1,U].$$ Functions $U$ and $U_1$ are
rational functions of the parameter $z$.\\

{\bf Remark 1.} A Baker function is determined up to an equivalence. Two Baker
functions $w^{(1)}$ and $w^{(2)}$ are equivalent if there are constant diagonal
matrices $c_k(z)=\sum_0^\infty c_{ki}z^{-i}=$diag $(c_{k\b}(z))$ such that
$w_k^{(1)}=w_k^{(2)}c_k$, i.e., $w_{k,\a\b}^{(1)}=c_{k,\b}w_{k,\a\b}^{(2)}$.
Equivalent Baker functions generate the same Lax operator $L$.\\

{\bf Remark 2.} Here we have a special case of ZS equation: the functions
$U$ and $U_1$ vanishing at infinity. If we make a gauge transformation $w_k
\mapsto g(t)w_k$ then $\p+U\mapsto g(\p+U)g^{-1}=\p+gUg^{-1}-
(\p g)g^{-1}$, the last term does not vanish at infinity. This yields
the general case.\\

If we deal with only one component $w_k$ of the Baker function, and consider
its
dependence solely on the variables $t_{k\a l}$ with the same $k$ (local
variables) ignoring all the others (alien variables), e.g., fixing their values
as parameters then we shall have a single-pole non-normalized hierarchy in the
sense of the previous section. (One has to perform a transformation $(z-a_k
)^{-1}=\z$). This fact allows to apply all the formulas obtained in that
section
to the present case. In particular, there are functions $\tau_k(t)$ and $\tau_{
k\a\b}(t)$ depending on local as well as on alien variables such that $$w_{k,\a
\b}(t,z)=c_{k\b}(z){G_{k\b}(z)\tau_{k,\a\b}(t)\over\tau_k(t)}e^{\xi_k}\eqno{(29
)}$$ where operators of translation $G_{k\b}(z)$ are defined by $$G_{k\b}f(t)
=f(...,t_{k_1,\g,l}-\d_{kk_1}\d_{\b\g}{1\over l}(z-a_k)^l,...)$$ and
$c_{k\b}(z)
$ are constant series in $z-a_k$.

We have not used yet the equations of the hierarchy with respect to the alien
variables. The rest of the section will be devoted to the proof that if those
equations are taken into account, then, roughly speaking, all denominators
$\tau_k$ in the previous formula are equal. More
precisely, the following theorem holds:\\

{\bf Theorem.} {\sl If $w=\{w_k\}$ is an arbitrary Baker function then
there are functions $\tau(t)$ and $\tau_{k,\a\b}(t)$ and
constant series $c_{k\b}(z)$ such that} $$w_{k,\a\b}(t,z)=c_{k\b}(z){G_{k\b}(z)
\tau_{k,\a\b}(t)\over\tau(t)}e^{\sum_l\xi_l}.$$

Notice that the last factor is $\exp\sum_l\xi_l$ and not just $\exp
\xi_k$, therefore the expression in front of it is not $\hw_k$. We call it
$\hhw_k$. Thus, $$w_k=\hhw_k\exp\sum_l\xi_l,~~\hw_k=\hhw_k\exi,~~
\hw_{k0}=\hhw_{k0}\exp\sum_{l\neq k}\xi_l(a_k).$$

{\em Proof.} We already have Baker functions $w_k$, $\hw_k=w_k\exp(-\xi_k)$ and
$\hhw_k=\hw_k\emxi$. Let us also introduce, as we did in Sect.6, $$v_k(z)=w
_{k0}^{-1}w_k(z),~~\hv_k(z)=v_k(z)\exp(-\xi_k)=w_{k0}^{-1}\hw_k(z)$$ and $$\hhv
_k(z)=\hhw_{k0}^{-1}\hhw_k(z)=\exp\sum_{l\neq k}\xi_l(a_k)\hv_k(z)
\emxi.$$ What
is important, $\hv_k$ and $\hhv_k$ differ by two diagonal factors, on the left
and on the right which do not depend on the variables with the same index $k$,
the local variables. The series $\hw_k$ and $\hhw_k$ differ by a right factor
of the same kind.

Considering $w_k$ as a function of local variables, we have noticed that this
is a Baker function of a single-pole not normalized hierarchy, and $v_k(z)$
that of the corresponding normalized hierarchy. Therefore, one can write for
them Eq.(22), or in present notations, $$G_{k\b}(\z)(\hv_k^{-1}(\z))_{\b\b}=(
\hv_{k,\b\b}(\z))^{-1}\eqno{(30)}$$ and Eq.(23), or
$$\hw_{k,\a\b}(\z)(\hv_{k,\b\b})^{-1}=G_{k\b}(\z)w_{k0,\a\b}.\eqno{(31)}$$ The
same equations can be written for
$\hhv_k$ and $\hhw_{k0}$ since the diagonal factors we discussed above will
cancel. They do not depend on local variables and the operators $G_{k\b}$ do
not act on them. Thus, $$G_{k\b}(\z)(\hhv_k^{-1}(\z))_{\b\b}=(\hhv_{k,\b\b}(\z)
)^{-1}\eqno{(30')}$$ and $$\hhw_{k,\a\b}(\z)(\hhv_{k,\b\b})^{-1}=G_{k\b}\hhw_{
k0,\a\b}.\eqno{(31')}$$ By the same reason we have the formula
$${G_{k\b_2}(\z_2)\hhv_{k,\b_1\b_1}(\z_1)\over\hhv_{k,\b_1\b_1}(\z_1
)}={G_{k\b_1}(\z_1)\hhv_{k,\b_2\b_2}(\z_2)\over\hhv_{k,\b_2\b_2}(\z_2)}.
\eqno{(32)}$$ It is correct for $\hv_k$ since this is, virtually, Eq.(5).
The additional diagonal factors cancel, so this is also correct for $\hhv_k$.\\

{\bf Lemma 1.} {\sl The equality $${G_{k_2\b}(\z_2)\hhv_{k_1,\b\b}(\z_1)\over
\hhv_{k_1,\b\b}(\z_1)}={G_{k_1\b}(\z_1)\hhv_{k_2,\b\b}(\z_2)\over\hhv_{k_2,\b\b
}(\z_2)}\eqno{(33)}$$ holds.} \\

{\em Proof of the lemma 1.} Eq.(27) implies that $\p_{k_1\a l}w_{k}\cdot w_{k}^
{-1}=B_{k_1\a l}=R_{k_1\a l}^-$ is a meromorphic function with a single pole at
$a_{k_1}$, vanishing at infinity and {\em not depending on} $k$. Actually, it
is easy to see that this is a characteristic property of the hierarchy which
expresses its universality, but we do not use this fact below. More generally,
$\p_{k_1\a_1 l}...\p_{k_s\a_s l}w_{k}\cdot w_k^{-1}$ does not depend on $k$ and
is a meromorphic function with the poles at $a_{k_1},...,a_{k_s}$ vanishing at
infinity when $s>0$. The same is also true for $w_k\cdot\p_{k_1\a_1 l}...\p_
{k_s\a_s l}w_{k}^{-1}$. This implies that the expression $(z-a_{k_1})^{-1}(z-a_
{k_2})^{-1}w_kG_{k_1\b}(\z_1)G_{k_2\b}(\z_2)w_k^{-1}$ is a meromorphic function
having the only poles on the Riemann sphere at $a_{k_1}$ and $a_{k_2}$ and not
depending on $k$. The sum of residues must vanish. Computing the residue at
$a_{k_1}$ we replace $k$ by $k_1$ and doing this at $a_{k_2}$ we replace $k$ by
$k_2$.
For simplicity of writing, let $\res_{k_i}$ symbolize $\res_{a_{k_i}}$. We have
$$\res_{k_1}(z-a_{k_1})^{-1}(z-a_{k_2})^{-1}w_{k_1}(z)G_{k_1\b}(\z_1)G_{k_2\b}(
\z_2)w_{k_1}^{-1}(z)$$ $$+\res_{k_2}(z-a_{k_1})^{-1}(z-a_{k_2})^{-1}w_{k_2}(z)
G_{k_1\b}(\z_1)G_{k_2\b}(\z_2)w_{k_2}^{-1}(z)=0.$$ In terms of $\hhw_k$ this
identity can be written as $$\res_{k_1}(z-a_{k_1})^{-1}(z-a_{k_2})^{-1}\hhw_{k_
1}(z)(I-E_\b+E_\b(1-{\z_1-a_{k_1}\over z-a_{k_1}})^{-1})$$ $$\cdot
(I-E_\b+E_\b(1-{\z_2-a_{k_2}\over z-a_{k_2}})^{-1})
G_{k_1\b}(\z_1)G_{k_2\b}(\z_2)\hhw_{k_1}^{-1}(z)+(k_1,\z_1\Leftrightarrow
k_2,\z_2)=0,$$ i.e.,$$\res_{k_1}(z-a_{k_1})^{-1}(z-a_{k_2})^{-1}\hhw_{k_1}(z)
(I-E_\b+E_\b(1-{\z_1-a_{k_1}\over z-a_{k_1}})^{-1}$$
$$\cdot(1-{\z_2-a_{k_2}\over z-a_{k_2}})^{-1})G_{k_1\b}(\z_1)G_{k_2\b}(\z_2)
\hhw_{k_1}^{-1}(z)+(k_1,\z_1\Leftrightarrow k_2,\z_2)=0,\eqno{(34)}$$
Here $(k_1,\z_1\Leftrightarrow k_2,\z_2)$ denotes a
term obtained by switching $k_1$ and $k_2$, $\z_1$ and $\z_2$. In the previous
sections we computed similar residues several times, so it does not need much
explanation. The term with $I-E_\b$ gives
$$(a_{k_1}-a_{k_2})\hhw_{k_10}(I-E_\b)
G_{k_1\b}(\z_1)G_{k_2\b}(\z_2)\hhw_{k_10}^{-1}+(k_1,\z_1\Leftrightarrow k_2,\z
_2).$$ Two others are
$$(\z_1-a_{k_2})^{-1}\hhw_{k_1}(\z_1)E_\b(1-{\z_2-a_{k_2}\over\z_1-a_{k_2}})^{-
1}G_{k_1\b}(\z_1)G_{k_2\b}(\z_2)\hhw_{k_1}^{-1}(\z_1)+(k_1,\z_1\Leftrightarrow
k_2,\z_2)$$ $$=(\z_1-\z_2)^{-1}[\hhw_{k_1}(\z_1)E_\b
G_{k_1\b}(\z_1)G_{k_2\b}(\z
_2)\hhw_{k_1}^{-1}(\z_1)-(k_1,\z_1\Leftrightarrow k_2,\z_2)].$$
Multiplying thus transformed Eq.(34) by $\hhw_{k_10}^{-1}$ on the left and by
$G_{k_1\b}(\z_1)G_{k_2\b}(\z_2)\hhw_{k_20}$ on the right we obtain $$(a_{k_1}-a
_{k_2})[(I-E_\b)
G_{k_1\b}(\z_1)G_{k_2\b}(\z_2)\hhw_{k_10}^{-1}\hhw_{k_20}-\hhw_{k_10}^{-1}\hhw
_{k_20}(I-E_\b)]$$ $$+(\z_1-\z_2)^{-1}[\hhv_{k_1}(\z_1)E_\b G_{k_1\b}(\z_1)
G_{k_2\b}(\z_2)\hhv_{k_1}^{-1}(\z_1)\hhw_{k_10}^{-1}\hhw_{k_20}$$ $$-
\hhw_{k_10}^{-1}\hhw_{k_20}\hhv_{k_2}(\z_2)E_\b G_{k_1\b}(\z_1)G_{k_2\b}(\z_2)
\hhv_{k_2}^{-1}(\z_2)]=0$$ where $\hhv_k(\z)=\hhw_{k0}^{-1}\hhw_k(\z)$. Now let
us take the $(\b,\b)$th element of this identity. The first two terms are not
involved in it, by virtue of the factors $I-E_\b$. Two others yield
$$\hhv_{k_1,
\b\b}G_{k_1\b}(\z_1)G_{k_2\b}(\z_2)(\hhv_{k_1}^{-1}(\z)T_{k_1k_2})_{\b\b}$$
$$=(T_{k_1k_2}\hhv_{k_2}(\z_2))_{\b\b}G_{k_1\b}(\z_1)G_{k_2\b}(\z_2)(\hhv_{k_2}^
{-1}(\z_2))_{\b\b}\eqno{(35)}$$ where $$T_{k_1k_2}=\hhw_{k_10}^{-1}\hhw_{k_20},
$$ the transition function. Using Eq.(30'), we can replace $G_{k_2\b}(\z_2)
(\hhv_{k_2}^{-1}(\z_2))_{\b\b}$ by $(\hhv_{k_2,\b\b}(\z_2))^{-1}$. Eq.(35)
becomes $$\hhv_{k,\b\b}G_{k_1\b}(\z_1)G_{k_2\b}(\z_2)(\hhv_{k_1}^{-1}(\z)T_{k_1
k_2})_{\b\b}$$ $$=(T_{k_1k_2}\hhv_{k_2}(\z_2))_{\b\b}G_{k_1\b}(\z_1)(\hhv_{k_2,
\b\b}(\z_2))^{-1}.\eqno{(36)}$$
Let $\z_1=a_{k_1}$. Then Eq.(36) becomes $$G_{k_2\b}(\z_2)T_{k_1k_2,\b\b}=
(T_{k_1k_2}\hhv_{k_2}(\z_2))_{\b\b}(\hhv_{k_2,\b\b}(\z_2))^{-1}.$$ This enables
us to rewrite the right-hand side of (36) as $$\hhv_{k_2,\b\b}(\z_2){
G_{k_2\b}(\z_2)T_{k_1k_2,\b\b}\over G_{k_1\b}(\z_1)\hhv_{k_2,\b\b}(\z_2)}.$$

Now, let $\z_2=a_{k_2}$. Eq.(36) transforms to $$\hhv_{k_1,\b\b}(\z_1)
G_{k_1\b}(\z_1)(\hhv_{k_1}^{-1}(\z)T_{k_1k_2})_{\b\b}=T_{k_1k_2,\b\b}.$$
The left-hand side of (36) can be written as
$$\hhv_{k_1,\b\b}(\z_1)G_{k_2\b}(\z
_2){T_{k_1k_2,\b\b}\over\hhv_{k_1,\b\b}(\z_1)}.$$ Equating the left- and the
right-hand sides and cancelling the common factor $G_{k_2\b}(\z_2)T_{k_1k_2,\b
\b}$, we obtain the required identity (33). $\Box$\\

{\bf Lemma 2.} {\sl The equation $${G_{k_2\b_2}(\z_2)\hhv_{k_1,\b_1\b_1}(\z_1)
\over
\hhv_{k_1,\b_1\b_1}(\z_1)}={G_{k_1\b_1}(\z_1)\hhv_{k_2,\b_2\b_2}(\z_2)\over\hhv
_{k_2,\b_2\b_2}(\z_2)}\eqno{(37)}$$ holds for any $k_1,k_2,\b_1$ and $\b_2$.}\\

{\em Proof of the lemma 2.} We already have two special cases of this lemma:
Eq.(32) for $k_1=k_2$ and lemma 1 for $\b_1=\b_2$. Now, suppose neither of
these conditions holds. Similarly to what we did proving the lemma 1, we write
a bilinear identity
$$\res_{k_1}(z-a_{k_1})^{-1}(z-a_{k_2})^{-1}w_{k_1}(z)G_{k_1\b_1}(\z_1)G_{k_2\b
_2}(\z_2)w_{k_1}^{-1}(z)+(k_1,\z_1,\b_1\Leftrightarrow k_2,\z_2,\b_2)=0.$$
In terms of $\hhw_k$ this
identity can be written as $$\res_{k_1}(z-a_{k_1})^{-1}(z-a_{k_2})^{-1}\hhw_{k_
1}(z)(I-E_{\b_1}+E_{\b_1}(1-{\z_1-a_{k_1}\over z-a_{k_1}})^{-1})$$ $$\cdot
(I-E_{\b_2}+E_{\b_2}(1-{\z_2-a_{k_2}\over z-a_{k_2}})^{-1})G_{k_1\b_1}(\z_1)G_{
k_2\b_2}(\z_2)\hhw_{k_1}^{-1}(z)+(k_1,\z_1,\b_1\Leftrightarrow k_2,\z_2,\b_2)=0
,$$ i.e.,$$\res_{k_1}(z-a_{k_1})^{-1}(z-a_{k_2})^{-1}\hhw_{k_1}(z)(I-E_{\b_1}-E
_{\b_2}+E_{\b_1}(1-{\z_1-a_{k_1}\over z-a_{k_1}})^{-1}$$ $$+E_{\b_2}(1-{\z_2-a_
{k_2}\over z-a_{k_2}})^{-1})G_{k_1\b_1}(\z_1)G_{k_2\b_2}(\z_2)\hhw_{k_1}^{-1}(z
)+(k_1,\z_1,\b_1\Leftrightarrow k_2,\z_2,\b_2)=0.$$ Computing the residues, we
have $$(a_{k_1}-a_{k_2})^{-1}\hhw_{k_10}(I-E_{\b_1}-E_{\b_2})
G_{k_1\b_1}(\z_1)G_{k_2\b_2}(\z_2)\hhw_{k_10}^{-1}$$ $$(\z_1-a_{k_2})^{-1}
\hhw_{k_1}(\z_1)E_{\b_1}G_{k_1\b_1}(\z_1)G_{k_2\b_2}(\z_2)\hhw_{k_1}^{-1}(\z_1)
$$ $$+(a_{k_1}-a_{k_2})^{-1}((1-{\z_2-a_{k_2}\over a_{k_1}-a_{k_2}})^{-1}\hhw_
{k_10}^{-1}+(k_1,\z_1,\b_1\Leftrightarrow k_2,\z_2,\b_2)=0.$$Dividing by $\hhw_
{k_10}$ on the left, by $G_{k_1\b_1}(\z_1)G_{k_2\b_2}(\z_2)\hhw_{k_20}^{-1}$ on
the right, we have $$*(I-E_{\b_1}-E_{\b_2})+(I-E_{\b_1}-E_{\b_2})*+(\z_1-a_{k_2
})^{-1}\hhv_{k_1}(\z_1)E_{\b_1}G_{k_1\b_1}(\z_1)G_{k_2\b_2}(\z_2)\hhv_{k_1}^{-1
}(\z_1)T_{k_1k_2}$$ $$+(\z_2-a_{k_1}^{-1}T_{k_1k_2}\hhv_{k_2}(\z_2)E_{\b_2}G_{k
_1\b_1}(\z_1)G_{k_2\b_2}(\z_2)\hv_{k_2}^{-1}-E_{\b_2}*+*E_{\b_1}=0$$ where
asterisks symbolize various factors which are not written in detail since they
are not important below.

Take the $(\b_1,\b_2)$th element of this equality. The terms with asterisks
vanish. The following terms remain: $$(\z_1-a_{k_2})^{-1}\hhv_{k_1\b_1\b_1}(\z_
1)G_{k_1\b_1}(\z_1)G_{k_2\b_2}(\z_2)(\hhv_{k_1}^{-1}(\z_1)T_{k_1k_2})_{\b_1\b_1}
$$
$$+(\z_2-a_{k_1})^{-1}(T_{k_1k_2}\hhv_{k_2}(\z_2))_{\b_1\b_2}G_{k_1\b_1}(\z_1
)G_{k_2\b_2}(\z_2)(\hv_{k_2}^{-1}(\z_2))_{\b_2\b_2}=0.\eqno{(38)}$$

Now, let $\z_1=a_{k_1}$: $$(a_{k_1}-a_{k_2})^{-1}G_{k_2\b_2}(\z_2)(T_{k_1k_2})
_{\b_1\b_2}$$ $$+(\z_2-a_{k_1})^{-1}(T_{k_1k_2}\hhv_{k_2}(\z_2))_{\b_1\b_2}
G_{k_2\b_2}(\z_2)(\hv_{k_2}^{-1})_{\b_2\b_2}=0.\eqno{(39)}$$Using this
equality, transform the second term of (38): $$-{(a_{k_1}-a_{k_2})^{-1}
G_{k_2\b_2}(\z_2)(T_{k_1k_2})_{\b_1\b_2}
G_{k_1\b_1}(\z_1)G_{k_2\b_2}(\z_2)(\hv_{k_2}^{-1}(\z_2))_{\b_2\b_2}
\over G_{k_2\b_2}(\z_2)(\hv_{k_2}^{-1})
_{\b_2\b_2}}$$ $$=-{(a_{k_1}-a_{k_2})^{-1}
G_{k_2\b_2}(\z_2)(T_{k_1k_2})_{\b_1\b_2}\hhv_{k_2,\b_2\b_2}(\z_2)\over
G_{k_1\b_1}(\z_1)\hhv_{k_2,\b_2\b_2}(\z_2)}.$$
(We have used Eq.(30') doing the last transformation).

Let $\z_2=a_{k_2}$:
$$ (\z_1-a_{k_2})^{-1}\hhv_{k_1\b_1\b_1}(\z_1)G_{k_1\b_1}(\z_1)(\hhv_{k_1}^{-1}
(\z_1)T_{k_1k_2})_{\b_1\b_1}+(a_{k_1}-a_{k_2})^{-1}(T_{k_1k_2})_{\b_1\b_2}=0
\eqno{(40)}$$ whence the first term can be written as $$-(a_{k_1}-a_{k_2})^{-1}
\hhv_{k_1\b_1\b_1}(\z_1)G_{k_2\b_2}(\z_2)\cdot{(T_{k_1k_2})_{\b_1\b_2}\over
\hhv_{k_1\b_1\b_1}(\z_1)}.$$
The identity (38) becomes, after a cancelation of the common factor,
$$ {\hhv_{k_2,\b_2\b_2}(\z_2)\over G_{k_1\b_1}(\z_1)\hhv_{k_2,\b_2\b_2}(\z_2)}=
{\hhv_{k_1\b_1\b_1}(\z_1)\over G_{k_2\b_2}(\z_2)\hhv_{k_1\b_1\b_1}(\z_1)}$$
which is the statement of the lemma. $\Box$    \\

{\bf Lemma 3.} {\sl The equation (37) implies that there is a function
$\tau(t)$ and constant series $c_{k\b}(\z)$ in powers of $\z-a_k$ such that
$\hhv_{k,\b\b}(\z)=c_{k\b}G_{k\b}\tau\cdot\tau^{-1}.$}  \\

{\em Proof of the lemma 3.} Taking the logarithm of (37) and denoting
$\ln\hhv_{k,\b\b}=f_{k,\b\b}$ we have $$(G_{k_1\b_1}(\z_1)-1)f_{k_2\b_2\b_2}
(\z_2)=(G_{k_2\b_2}(\z_2)-1)f_{k_1\b_1\b_1}(\z_1),$$ and we just have to repeat
the derivation of the Eq.(8) from (5) in Sect.4. $\Box$

{\em The end of the proof of the theorem.} Put $\tau_{k,\a\b}=\tau\cdot\hhw_{k0
\a\b}$. Taking into account (31'), we have $$ {G_{k\b}(\z)\tau_{k,\a\b}\over
\tau}={G_{k\b}(\z)\tau\over\tau}\cdot G_{k\b}(\z)\hhw_{k0,\a\b}=c_{k\b}^{-1}
\hhv_{k,\b\b}\cdot\hhw_{k,\a\b}\hhv_{k,\b\b}^{-1}=c_{k\b}^{-1}\hhw_{k,\a\b}$$
as required. $\Box$    \\

{\bf References.}\\

\noindent 1. Sato, M.:  Soliton equations as dynamical systems on infinite
dimensional Grassmann manifolds, RIMS Kokyuroku, 439, 30-46, 1981.\\

\noindent 2. Date, E., Jimbo, M., Kashiwara, M., and Miwa, T.: Transformation
groups for soliton equations, in: Jimbo and Miwa (ed.) Non-linear integrable
systems -- classical theory and quantum theory, Proc. RIMS symposium,
Singapore,
1983.\\

\noindent 3. Date, E., Jimbo, M., Kashiwara, M., and Miwa, T.: Operator
approach to the Kadomtsev-Petviashvili equation - transformation groups for
soliton equations III, Journ. Phys. Soc. Japan, 50, 3806-3812, 1981.\\

\noindent 4. Ueno, K., and Takasaki, K.: Toda lattice hierarchy, in: Advanced
Studies in Pure Mathematics, 4, World Scientific, 1-95, 1984.\\

\noindent 5. Dickey, L. A.: On Segal-Wilson's definition of the $\tau$-function
and hierarchies AKNS-D and mcKP, in: Integrable systems, The Verdier Memorial
Conference, Birkh\"auser, 147-162, 1993.\\

\noindent 6. Dickey, L. A.: On the $\t$-function of matrix hierarchies of
integrable equations, Journal Math. Physics, 32, 2996-3002, 1991.\\

\noindent 7. Dickey, L. A.: Why the general Zakharov-Shabat equations form a
hierarchy, Com. Math. Phys., 163, 509-521, 1994.\\

\noindent 8. Vasilev, S.: Tau functions of algebraic geometrical solutions to
the general Zakharov-Shabat hierarchy, Preprint of the University of Oklahoma,
1994.\\

\noindent 9. Dickey, L. A.: Soliton Equations and Hamiltonian Systems, Advanced
Series in Mathematical Physics, 12, World Scientific, 1991.\\
\end{document}